\documentclass[12pt,a4paper]{article}

\usepackage{amsmath}
\usepackage{ifthen} 
\newboolean{pdflatex}
\setboolean{pdflatex}{true} 

\usepackage{float}
\floatstyle{plaintop}
\restylefloat{table}

\usepackage{graphicx}
\usepackage{subfig}
\usepackage{multirow}

\usepackage{caption}

\newboolean{articletitles}
\setboolean{articletitles}{true} 

\newboolean{uprightparticles}
\setboolean{uprightparticles}{false} 

\usepackage{microtype}
\usepackage{lineno}  
\usepackage{xspace} 

\usepackage{graphicx}  
\usepackage{color}
\usepackage{colortbl}
\graphicspath{{./figs/}} 

\usepackage{amsmath} 
\usepackage{amssymb}
\usepackage{amsfonts}
\usepackage{upgreek} 

\newcommand*\patchAmsMathEnvironmentForLineno[1]{%
\expandafter\let\csname old#1\expandafter\endcsname\csname #1\endcsname
\expandafter\let\csname oldend#1\expandafter\endcsname\csname
end#1\endcsname
 \renewenvironment{#1}%
   {\linenomath\csname old#1\endcsname}%
   {\csname oldend#1\endcsname\endlinenomath}%
}
\newcommand*\patchBothAmsMathEnvironmentsForLineno[1]{%
  \patchAmsMathEnvironmentForLineno{#1}%
  \patchAmsMathEnvironmentForLineno{#1*}%
}
\AtBeginDocument{%
\patchBothAmsMathEnvironmentsForLineno{equation}%
\patchBothAmsMathEnvironmentsForLineno{align}%
\patchBothAmsMathEnvironmentsForLineno{flalign}%
\patchBothAmsMathEnvironmentsForLineno{alignat}%
\patchBothAmsMathEnvironmentsForLineno{gather}%
\patchBothAmsMathEnvironmentsForLineno{multline}%
}

\usepackage{hyperref}    
\usepackage[all]{hypcap} 

\def\lhcb {\mbox{LHCb}\xspace}
\def\ux85 {\mbox{UX85}\xspace}

\ifthenelse{\boolean{uprightparticles}}%
{

 \def\Ppi         {\ensuremath{\uppi}\xspace}

 \def\PDelta      {\ensuremath{\Delta}\xspace}                 
 \def\PXi      {\ensuremath{\Xi}\xspace}                 
 \def\PLambda      {\ensuremath{\Lambda}\xspace}                 
 \def\PSigma      {\ensuremath{\Sigma}\xspace}                 
 \def\POmega      {\ensuremath{\Omega}\xspace}                 
 \def\PUpsilon      {\ensuremath{\Upsilon}\xspace}                 
 
 \def\PB      {\ensuremath{\mathrm{B}}\xspace}                 
                  
 \def\PD      {\ensuremath{\mathrm{D}}\xspace}

 \def\PK      {\ensuremath{\mathrm{K}}\xspace}

 \def\Pb      {\ensuremath{\mathrm{b}}\xspace}                 
 \def\Pc      {\ensuremath{\mathrm{c}}\xspace}

 \def\Pi      {\ensuremath{\mathrm{i}}\xspace}

}
{

 \def\Ppi         {\ensuremath{\pi}\xspace}

 \mathchardef\PDelta="7101
 \mathchardef\PXi="7104
 \mathchardef\PLambda="7103
 \mathchardef\PSigma="7106
 \mathchardef\POmega="710A
 \mathchardef\PUpsilon="7107
                  
 \def\PB      {\ensuremath{B}\xspace}                 
                  
 \def\PD      {\ensuremath{D}\xspace}

 \def\PK      {\ensuremath{K}\xspace}

 \def\Pb      {\ensuremath{b}\xspace}                 
 \def\Pc      {\ensuremath{c}\xspace}

 \def\Pi      {\ensuremath{i}\xspace}

}

\def\cquark    {\ensuremath{\Pc}\xspace}

\def\bquark    {\ensuremath{\Pb}\xspace}

\def\pion  {\ensuremath{\Ppi}\xspace}

\def\pip   {\ensuremath{\pion^+}\xspace}
\def\pim   {\ensuremath{\pion^-}\xspace}

\def\kaon  {\ensuremath{\PK}\xspace}
  \def\Kbar  {\kern 0.2em\overline{\kern -0.2em \PK}{}\xspace}

\def\Kz    {\ensuremath{\kaon^0}\xspace}
\def\Kzb   {\ensuremath{\Kbar^0}\xspace}
\def\KzKzb {\ensuremath{\Kz \kern -0.16em \Kzb}\xspace}
\def\Kp    {\ensuremath{\kaon^+}\xspace}
\def\Km    {\ensuremath{\kaon^-}\xspace}

\def\KpKm  {\ensuremath{\Kp \kern -0.16em \Km}\xspace}

\def\Kstarzb {\ensuremath{\Kbar^{*0}}\xspace}

  \def\Dbar    {\kern 0.2em\overline{\kern -0.2em \PD}{}\xspace}
\def\D       {\ensuremath{\PD}\xspace}

\def\Dz      {\ensuremath{\D^0}\xspace}
\def\Dzb     {\ensuremath{\Dbar^0}\xspace}
\def\DzDzb   {\ensuremath{\Dz {\kern -0.16em \Dzb}}\xspace}
\def\Dp      {\ensuremath{\D^+}\xspace}
\def\Dm      {\ensuremath{\D^-}\xspace}

\def\DpDm    {\ensuremath{\Dp {\kern -0.16em \Dm}}\xspace}

\def\Dstarm  {\ensuremath{\D^{*-}}\xspace}

\def\B       {\ensuremath{\PB}\xspace}
\def\Bbar    {\ensuremath{\kern 0.18em\overline{\kern -0.18em \PB}{}}\xspace}

\def\Bz      {\ensuremath{\B^0}\xspace}

  \def\Y#1S{\ensuremath{\PUpsilon{(#1S)}}\xspace}

\def\Lbar {\ensuremath{\kern 0.1em\overline{\kern -0.1em\PLambda}}\xspace}

\def\to                 {\ensuremath{\rightarrow}\xspace}

\def\CP                {\ensuremath{C\!P}\xspace}

\def\AT#1     {\ensuremath{A_{\mathrm{T}}^{#1}}\xspace}           

\def\C#1      {\ensuremath{\mathcal{C}_{#1}}\xspace}                       
\def\Cp#1     {\ensuremath{\mathcal{C}_{#1}^{'}}\xspace}                    
\def\Ceff#1   {\ensuremath{\mathcal{C}_{#1}^{\mathrm{(eff)}}}\xspace}        
\def\Cpeff#1  {\ensuremath{\mathcal{C}_{#1}^{'\mathrm{(eff)}}}\xspace}       
\def\Ope#1    {\ensuremath{\mathcal{O}_{#1}}\xspace}                       
\def\Opep#1   {\ensuremath{\mathcal{O}_{#1}^{'}}\xspace}                    

\newcommand{\tev}{\ensuremath{\mathrm{\,Te\kern -0.1em V}}\xspace}
\newcommand{\gev}{\ensuremath{\mathrm{\,Ge\kern -0.1em V}}\xspace}
\newcommand{\mev}{\ensuremath{\mathrm{\,Me\kern -0.1em V}}\xspace}
\newcommand{\kev}{\ensuremath{\mathrm{\,ke\kern -0.1em V}}\xspace}
\newcommand{\ev}{\ensuremath{\mathrm{\,e\kern -0.1em V}}\xspace}
\newcommand{\gevc}{\ensuremath{{\mathrm{\,Ge\kern -0.1em V\!/}c}}\xspace}
\newcommand{\mevc}{\ensuremath{{\mathrm{\,Me\kern -0.1em V\!/}c}}\xspace}
\newcommand{\gevcc}{\ensuremath{{\mathrm{\,Ge\kern -0.1em V\!/}c^2}}\xspace}
\newcommand{\gevgevcccc}{\ensuremath{{\mathrm{\,Ge\kern -0.1em V^2\!/}c^4}}\xspace}
\newcommand{\mevcc}{\ensuremath{{\mathrm{\,Me\kern -0.1em V\!/}c^2}}\xspace}

\def\mum  {\ensuremath{\,\upmu\rm m}\xspace}

\newcommand{\chisq}{\ensuremath{\chi^2}\xspace}

\def\gsim{{~\raise.15em\hbox{$>$}\kern-.85em
          \lower.35em\hbox{$\sim$}~}\xspace}
\def\lsim{{~\raise.15em\hbox{$<$}\kern-.85em
          \lower.35em\hbox{$\sim$}~}\xspace}

\def\pt         {\mbox{$p_{\rm T}$}\xspace}

\def\evtgen     {\mbox{\textsc{EvtGen}}\xspace}
\def\pythia     {\mbox{\textsc{Pythia}}\xspace}

\def\geant      {\mbox{\textsc{Geant4}}\xspace}

\def\photos     {\mbox{\textsc{Photos}}\xspace}

\def\tell1  {TELL1\xspace}
\def\ukl1   {UKL1\xspace}

\usepackage{cite} 
\usepackage{mciteplus}

\usepackage{longtable} 

\begin{document}

\renewcommand{\thefootnote}{\fnsymbol{footnote}}
\setcounter{footnote}{1}


\begin{titlepage}
\pagenumbering{roman}

\vspace*{-1.5cm}
\centerline{\large EUROPEAN ORGANIZATION FOR NUCLEAR RESEARCH (CERN)}
\vspace*{1.5cm}
\hspace*{-0.5cm}
\begin{tabular*}{\linewidth}{lc@{\extracolsep{\fill}}r}
\ifthenelse{\boolean{pdflatex}}
{\vspace*{-2.7cm}\mbox{\!\!\!\includegraphics[width=.14\textwidth]{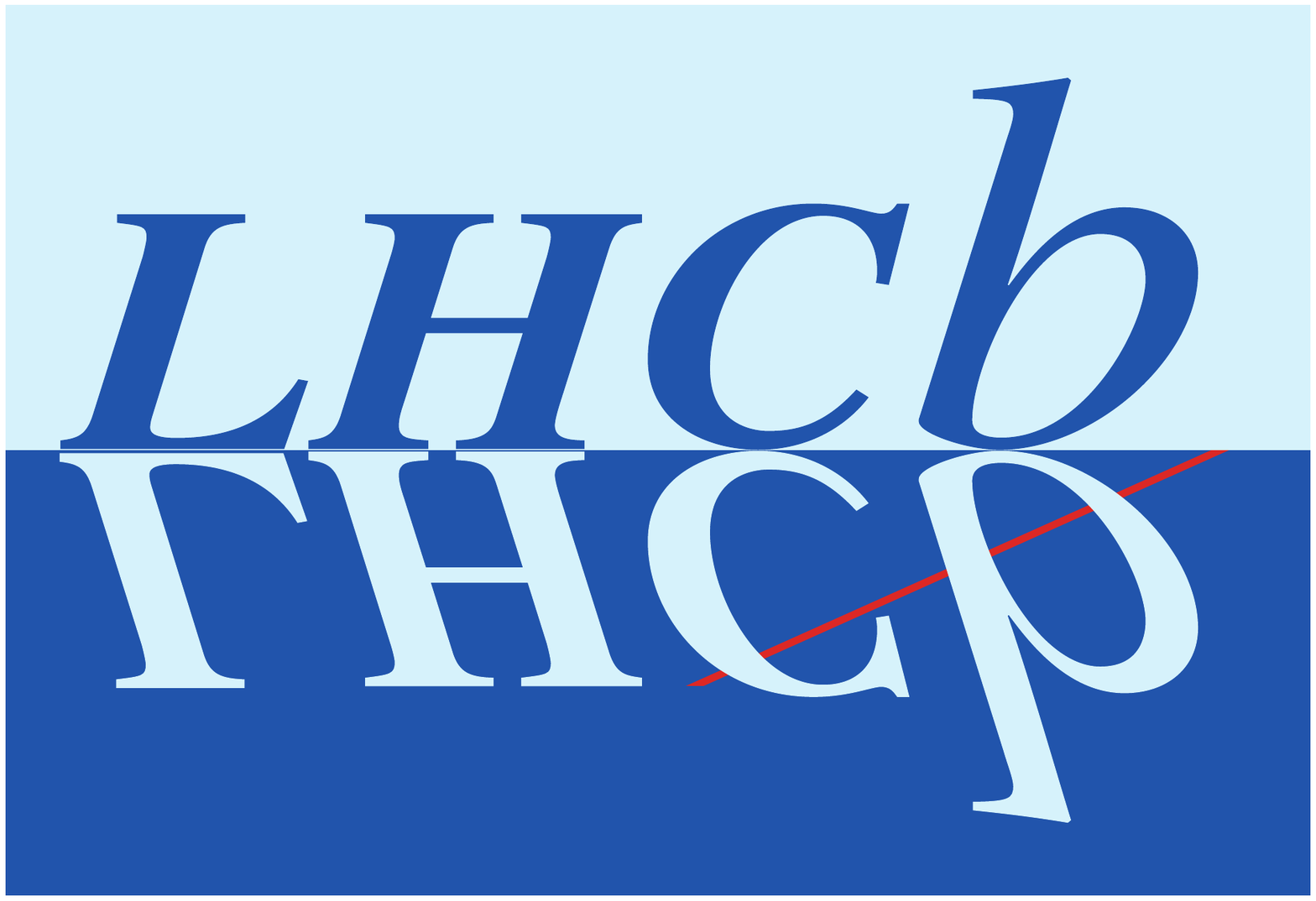}} & &}%
{\vspace*{-1.2cm}\mbox{\!\!\!\includegraphics[width=.12\textwidth]{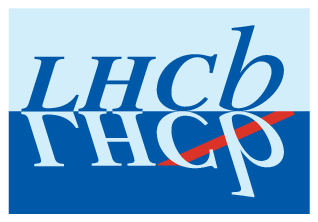}} & &}%
\\
 & & CERN-PH-EP-2013-041 \\  
 & & LHCb-PAPER-2012-046 \\  
 & & 27 March, 2013 \\ 
 & & \\
\end{tabular*}

\vspace*{1.3cm}

{\bf\boldmath\huge
\begin{center}
Study of $B^{0} \rightarrow D^{*-}\pi^{+}\pi^{-}\pi^{+}$ and $B^{0} \rightarrow D^{*-}K^{+}\pi^{-}\pi^{+}$ decays
\end{center}
}

\vspace*{0.8cm}

\begin{center}
The LHCb collaboration\footnote{Authors are listed on the following pages.}
\end{center}


\begin{abstract}
  \noindent
Using proton-proton collision data collected by the LHCb experiment at \mbox{$\sqrt{s} =$ 7 TeV}, corresponding to an integrated luminosity of 1.0 fb$^{-1}$, the ratio of branching fractions of the $B^{0}\rightarrow D^{*-}\pi^{+}\pi^{-}\pi^{+}$ decay relative to the $B^{0} \rightarrow D^{*-}\pi^{+}$ decay is measured to be
\begin{equation*}
\frac{\mathcal{B}(B^{0}\rightarrow D^{*-}\pi^{+}\pi^{-}\pi^{+})}{\mathcal{B}(B^{0} \rightarrow D^{*-}\pi^{+})} = 2.64 \pm 0.04\,(\text{stat.}) \pm 0.13\,(\text{syst.})\, .
\label{eq:CF_BF_ratio_result}
\end{equation*}
The Cabibbo-suppressed decay $B^{0}\rightarrow D^{*-}K^{+}\pi^{-}\pi^{+}$ is observed for the first time and the measured ratio of branching fractions is
\begin{equation*}
\frac{\mathcal{B}(B^{0}\rightarrow D^{*-}K^{+}\pi^{-}\pi^{+})}{\mathcal{B}(B^{0} \rightarrow D^{*-}\pi^{+}\pi^{-}\pi^{+})} = (6.47 \pm 0.37\,(\text{stat.}) \pm 0.35\,(\text{syst.})) \times 10^{-2}
\label{eq:CS_BF_ratio_result}\, .
\end{equation*}
A search for orbital excitations of charm mesons contributing to the \mbox{$B^{0}\rightarrow D^{*-}\pi^{+}\pi^{-}\pi^{+}$} final state is also performed, and the first observation of the \mbox{$B^{0}\rightarrow \Dbar_{1}(2420)^{0}\pi^{+}\pi^{-}$} decay is reported with the ratio of branching fractions
\begin{equation*}
\frac{\mathcal{B}(B^{0}\rightarrow (\Dbar_{1}(2420)^{0} \to D^{*-} \pi^{+}) \pi^{-}\pi^{+})}{\mathcal{B}(B^{0} \rightarrow D^{*-}\pi^{+}\pi^{-}\pi^{+})} = (2.04 \pm 0.42\,(\text{stat.}) \pm 0.22\,(\text{syst.})) \times 10^{-2}\, ,
\end{equation*}
where the numerator represents a product of the branching fractions \mbox{$\mathcal{B}(B^{0}\rightarrow \Dbar_{1}(2420)^{0} \pim \pip)$} and \mbox{$\mathcal{B}(\Dbar_{1}(2420)^{0} \to D^{*-} \pi^{+})$}.
\end{abstract}

\vspace*{1.5cm}

\begin{center}
  Submitted to Phys. Rev. D
\end{center}

\vspace{\fill}

{\footnotesize 
\centerline{\copyright~CERN on behalf of the \lhcb collaboration, license \href{http://creativecommons.org/licenses/by/3.0/}{CC-BY-3.0}.}}
\vspace*{2mm}

\end{titlepage}



\centerline{\large\bf LHCb collaboration}
\begin{flushleft}
\small
R.~Aaij$^{38}$, 
C.~Abellan~Beteta$^{33,n}$, 
A.~Adametz$^{11}$, 
B.~Adeva$^{34}$, 
M.~Adinolfi$^{43}$, 
C.~Adrover$^{6}$, 
A.~Affolder$^{49}$, 
Z.~Ajaltouni$^{5}$, 
J.~Albrecht$^{9}$, 
F.~Alessio$^{35}$
% ...

{\footnotesize \it
% ...
$ ^{5}$Clermont Universit\'{e}, Universit\'{e} Blaise Pascal, CNRS/IN2P3, LPC, Clermont-Ferrand, France\\
$ ^{6}$CPPM, Aix-Marseille Universit\'{e}, CNRS/IN2P3, Marseille, France\\
% ...
$ ^{9}$Fakult\"{a}t Physik, Technische Universit\"{a}t Dortmund, Dortmund, Germany\\
% ...
$ ^{11}$Physikalisches Institut, Ruprecht-Karls-Universit\"{a}t Heidelberg, Heidelberg, Germany\\
% ...
$ ^{33}$Universitat de Barcelona, Barcelona, Spain\\
$ ^{34}$Universidad de Santiago de Compostela, Santiago de Compostela, Spain\\
$ ^{35}$European Organization for Nuclear Research (CERN), Geneva, Switzerland\\
% ...
$ ^{38}$Nikhef National Institute for Subatomic Physics, Amsterdam, The Netherlands\\
% ...
$ ^{43}$H.H. Wills Physics Laboratory, University of Bristol, Bristol, United Kingdom\\
% ...
$ ^{49}$Oliver Lodge Laboratory, University of Liverpool, Liverpool, United Kingdom\\
% ...
$ ^{n}$LIFAELS, La Salle, Universitat Ramon Llull, Barcelona, Spain\\
% ...
}
\end{flushleft}

\cleardoublepage


\renewcommand{\thefootnote}{\arabic{footnote}}
\setcounter{footnote}{0}



\pagestyle{plain} 
\setcounter{page}{1}
\pagenumbering{arabic}



%

\section{Introduction}

Open charm decays of $b$ hadrons offer a means by which both the electroweak and QCD sectors of the Standard Model (SM) may be tested.
Beyond measurements of \CP violation and the phases derived from the CKM matrix, rare $B\to DX$ decays may be used to search for new physics in decays mediated via annihilation or exchange processes. 
High multiplicity $B\to D X$ decays are receiving increasing attention at LHCb~\cite{LHCb-PAPER-2011-016,LHCb-PAPER-2011-040,LHCb-PAPER-2012-033}, in part owing to the large samples that can be obtained as a result of the copious $b\overline{b}$ production at the LHC~\cite{production_CS}.

Improving knowledge of the $B^{0}\to D^{(*)-}\pi^{+}\pi^{-}\pi^{+}$ decay is of interest because of its potential use as a normalisation mode for the semileptonic decay $B^{0} \rightarrow D^{(*)-}\tau^{+}\nu_{\tau}$ with $\tau^{+}\rightarrow \pi^{+}\pi^{-}\pi^{+}\overline{\nu}_{\tau}$~\cite{Anne_thesis}. Charge conjugation is implied throughout. The latter $B$ decay has recently shown an excess over the SM branching fraction expectation~\cite{Lees:2012xj}, which could indicate the presence of physics beyond the SM.

In this work three measurements of ratios of branching fractions are described.  The first measurement is of the Cabibbo-favoured (CF) ratio
\begin{equation*}
r_{3h} = \frac{\mathcal{B}(B^{0}\rightarrow D^{*-}\pi^{+}\pi^{-}\pi^{+})}{\mathcal{B}(B^{0}\rightarrow D^{*-}\pi^{+})}\, ,
\end{equation*}
from which a value for $\mathcal{B}(B^{0}\rightarrow D^{*-}\pi^{+}\pi^{-}\pi^{+})$ is obtained. The current world average value is $\mathcal{B}(B^{0}\rightarrow D^{*-}\pi^{+}\pi^{-}\pi^{+}) =$ (7.0 $\pm$ 0.8) $\times 10^{-3}$~\cite{PDG2012}. The subscript $3h$ denotes the three hadrons (or ``bachelors") in the signal $B^{0}$ decay, which are produced along with the charmed meson. The second measurement is of the ratio of branching fractions of the Cabibbo-suppressed (CS) decay $B^{0}\rightarrow D^{*-}K^{+}\pi^{-}\pi^{+}$ relative to its CF counterpart
\begin{equation*}
r_{\text{CS}}^{K\pi\pi} = \frac{\mathcal{B}(B^{0}\rightarrow D^{*-}K^{+}\pi^{-}\pi^{+})}{\mathcal{B}(B^{0}\rightarrow D^{*-}\pi^{+}\pi^{-}\pi^{+})}\, ,
\end{equation*}
where the subscript denotes Cabibbo suppression of the signal decay. The third measurement is of the ratio of branching fractions of the CS decay $B^{0}\rightarrow D^{*-}K^{+}$ relative to its CF counterpart
\begin{equation*}
r_{\text{CS}}^{K} = \frac{\mathcal{B}(B^{0}\rightarrow D^{*-}K^{+})}{\mathcal{B}(B^{0}\rightarrow D^{*-}\pi^{+})}\,.
\end{equation*}
The superscripts in $r_{\text{CS}}^{K}$ and $r_{\text{CS}}^{K\pi\pi}$ are used to differentiate the single- and triple-bachelor measurements, respectively. The rate of the CS processes are expected to be smaller than their CF counterparts by a factor of roughly tan$^{2}(\theta_{\text{C}}) \sim 1/20$, where $\theta_{\text{C}}$ is the Cabibbo angle~\cite{PDG2012}.

\section{Detector, signal selection and simulation}

The \lhcb detector~\cite{Alves:2008zz} is a single-arm forward spectrometer covering the \mbox{pseudorapidity} range $2<\eta <5$, designed for the study of particles containing \bquark or \cquark quarks. The detector includes a high precision tracking system consisting of a silicon-strip vertex detector surrounding the $pp$ interaction region, a large-area silicon-strip detector located upstream of a dipole magnet with a bending power of about $4{\rm\,Tm}$, and three stations of silicon-strip detectors and straw drift tubes placed downstream. The combined tracking system has a momentum resolution $\Delta p/p$ that varies from 0.4\% at 5\gevc to 0.6\% at 100\gevc, and an impact parameter (IP) resolution of 20\mum for tracks with high \mbox{transverse momentum (\pt)}. Charged hadrons are identified using two ring-imaging Cherenkov (RICH) detectors, which provide a kaon identification efficiency of $\sim$95\% for a pion fake rate of a few percent, integrated over the momentum range from 3--100 GeV/$c$~\cite{arXiv:1211-6759}. Photon, electron and hadron candidates are identified by a calorimeter system consisting of scintillating-pad and preshower detectors, an electromagnetic calorimeter and a hadronic calorimeter. Muons are identified by a system composed of alternating layers of iron and multiwire proportional chambers. 

LHCb uses a two-level trigger system. The first level of the trigger consists of a hardware stage that searches for either a large transverse energy cluster ($E_{\text{T}} >$ 3.6 GeV) in the calorimeters, or a single high $\pt$ muon or di-muon pair in the muon stations. This is followed by a software stage that applies a full event reconstruction. The software trigger requires a two-, three- or four-track secondary vertex with a high sum of the \pt of the tracks and a significant displacement from the primary $pp$ interaction vertices. At least one track should have $\pt > 1.7\gevc$ and \mbox{IP \chisq} with respect to the primary interaction greater than 16. The IP \chisq is defined as the difference between the \chisq of the primary vertex (PV) reconstructed with and without the considered track. A multivariate algorithm is used for the identification of secondary vertices consistent with the decay of a \bquark hadron~\cite{Aaij:2012me}. The results presented in this paper use the full 2011 dataset corresponding to an integrated luminosity of 1.0 fb$^{-1}$ of $pp$ collisions at \mbox{$\sqrt{s} =$ 7 TeV}.

Candidate $B^{0}$ decays are formed by combining a $D^{*-} \rightarrow (\Dzb \rightarrow K^{+}\pi^{-})\pi^{-}$ candidate with either a single bachelor or triple-bachelor combination. All final state tracks are required to have \pt in excess of 500 MeV/$c^{2}$, except for the slow pion produced in the $D^{*-}$ decay, which must have \pt in excess of 100 \mevc. All tracks must be well separated from any reconstructed PV in the event. They must be identified as either a pion or a kaon using information from the RICH detectors. Particle identification likelihoods for several hypotheses (e.g. $\pip, K^{+}, p$) are formed, and the difference in the logarithms of these likelihoods, $\Delta \text{LL}$, is used to differentiate between particle types.  Candidate \Dzb mesons are required to have good vertex fit quality, be well displaced from the nearest PV and have an invariant mass $m(K^{+}\pi^{-}$) within 50 MeV/$c^{2}$ of the \Dzb mass. Candidate $D^{*-}$ decays are selected by requiring \mbox{140 $< m(K^{+}\pi^{-}\pi^{+}) - m(K^{+}\pi^{-}) <$ 150 MeV/$c^{2}$}. 

Candidate triple-bachelors are formed from a $\pi^{+}\pi^{-}\pi^{+}$ or $K^{+}\pi^{-}\pi^{+}$ combination, where all invariant mass values up to 3 GeV/$c^{2}$ are accepted. The vertex of the combination must be well separated from the nearest PV. Backgrounds from $B^{0} \rightarrow D^{*-} \pi^{+}(\pi^{-}\pi^{+})$ decays for the CS modes $B^{0} \rightarrow D^{*-} K^{+}(\pi^{-}\pi^{+})$ are reduced by applying more stringent particle identification (PID) requirements to the bachelor kaon. To suppress backgrounds from $B^{0} \rightarrow D^{*-} D_{s}^{+}$ decays where $D_{s}^{+} \rightarrow K^{+}\pi^{-}\pi^{+}$ in the triple-bachelor decay \mbox{$B^{0} \rightarrow D^{*-} K^{+}\pi^{-}\pi^{+}$}, it is required that $m(K^{+}\pi^{-}\pi^{+})$ is more than 15 MeV/$c^{2}$ away from the $D_{s}^{+}$ mass. Reconstructed $B^{0}$ candidates are required to be well separated from the nearest PV, with decay time larger than 0.2 ps and good quality vertex fit. Candidates passing all selection requirements are refit with both \Dzb mass and vertex constraints to improve the $B^{0}$ mass resolution~\cite{Hulsbergen:2005pu}.

Selection efficiencies and trigger pass fractions defined below are evaluated using Monte Carlo simulation. In the simulation, $pp$ collisions are generated using \pythia~6.4~\cite{Sjostrand:2006za} with a specific \lhcb
configuration~\cite{LHCb-PROC-2010-056}.  Decays of hadronic particles are described by \evtgen~\cite{Lange:2001uf} in which final state
radiation is generated using \photos~\cite{Golonka:2005pn}. The
interaction of the generated particles with the detector and its
response are implemented using the \geant
toolkit~\cite{Allison:2006ve, *Agostinelli:2002hh} as described in
Ref.~\cite{LHCb-PROC-2011-006}.

The simulated events are passed through an emulation of the hardware trigger and then through the full software trigger as run on data.
The total kinematic efficiency, $\epsilon_{\text{kin}}$, is determined from the simulation as the fraction of events that pass all reconstruction and selection requirements and the trigger. The fraction of selected events that pass the particular trigger selection relative to the total number of selected events is taken to be the trigger pass fraction, $f_{\text{trig}}$. This fraction does not represent the true trigger efficiency, since it is evaluated with respect to a sample of events that have all passed the trigger.

\section{Fits to data}

\begin{figure}[!htpb]
\begin{center}
\subfloat{\label{fig:KPiPi_Fit}\includegraphics[width=0.5\textwidth]{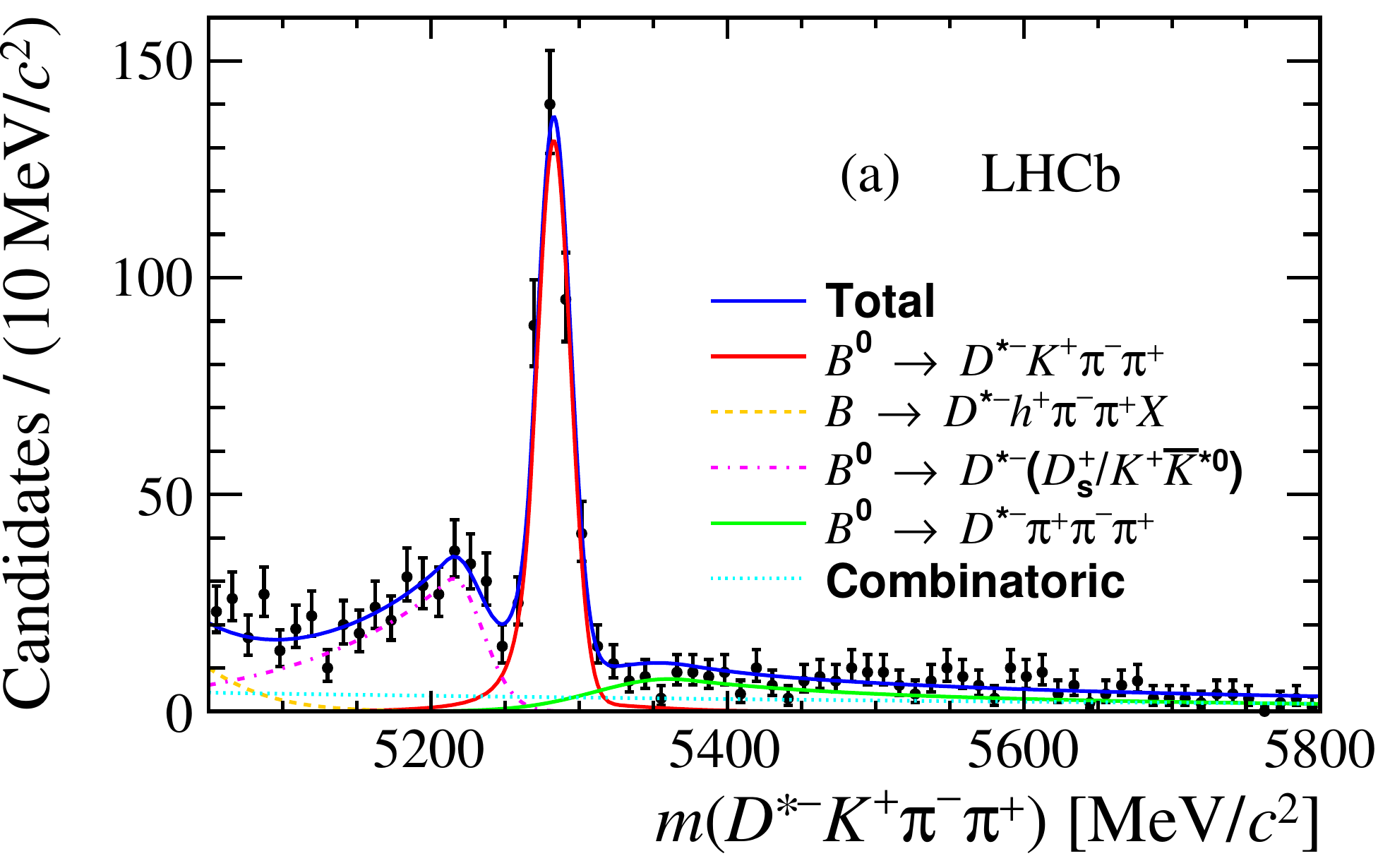}}
\subfloat{\label{fig:K_Fit}\includegraphics[width=0.5\textwidth]{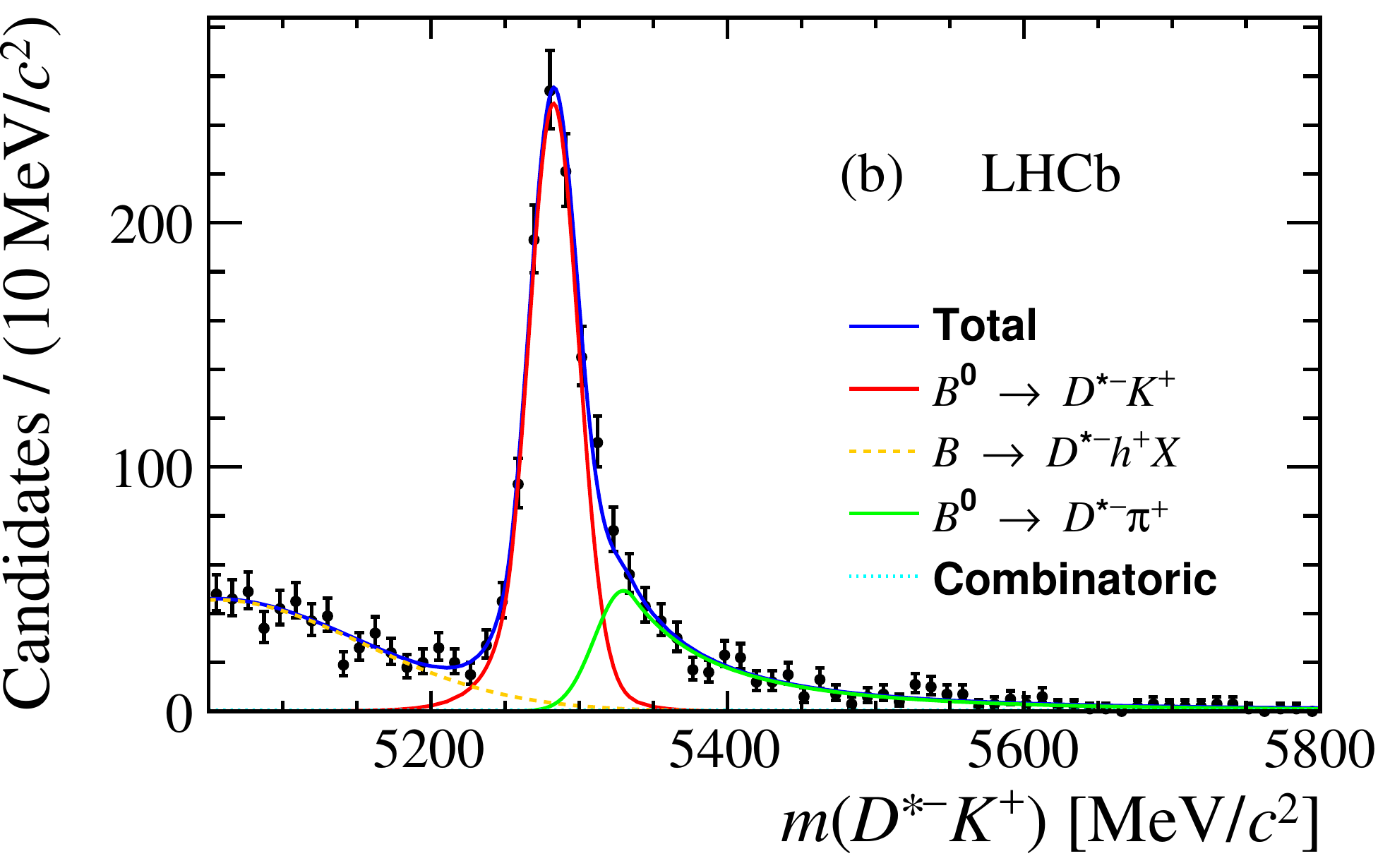}}\\
\subfloat{\label{fig:PiPiPi_Fit}\includegraphics[width=0.5\textwidth]{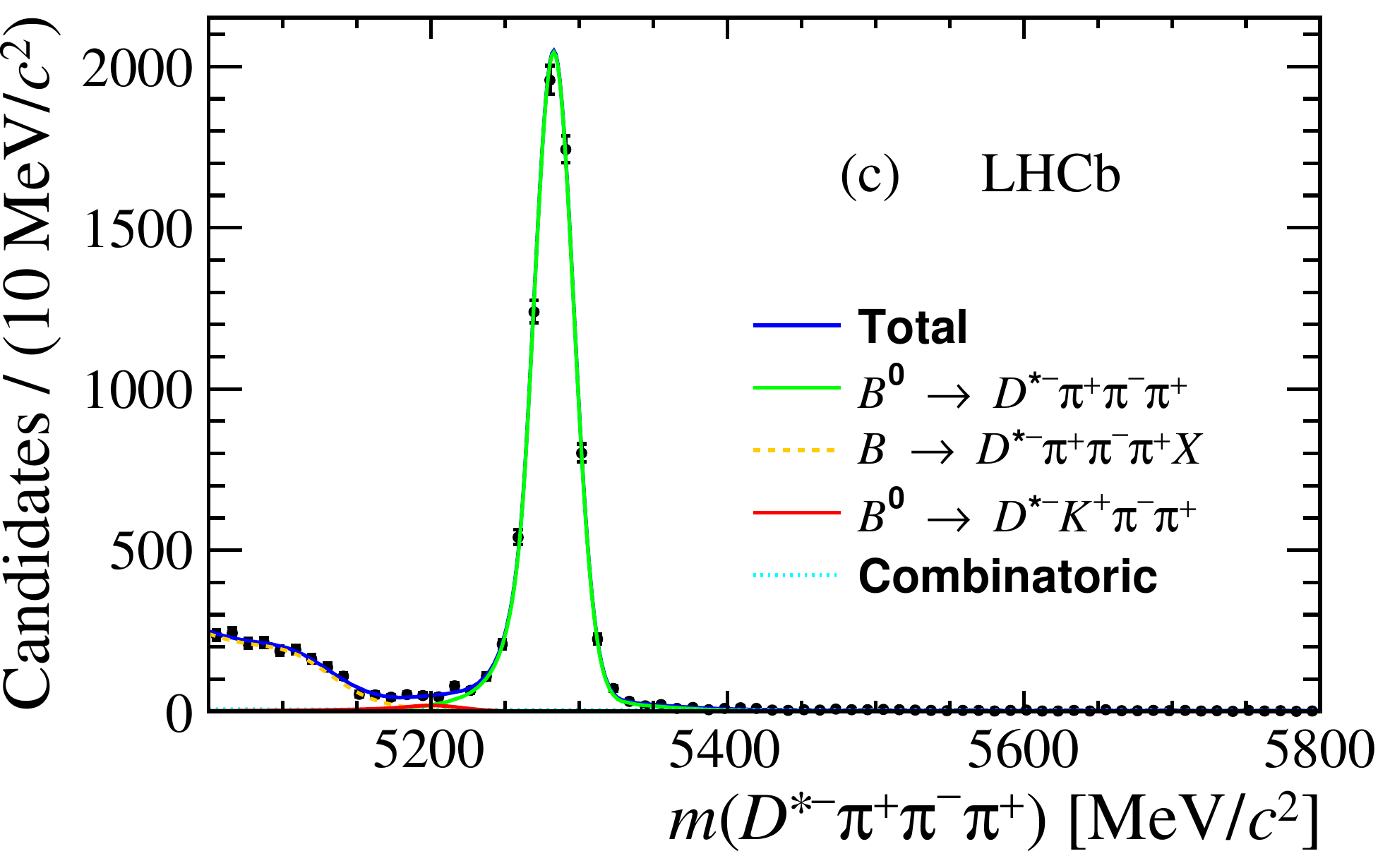}}
\subfloat{\label{fig:Pi_Fit}\includegraphics[width=0.5\textwidth]{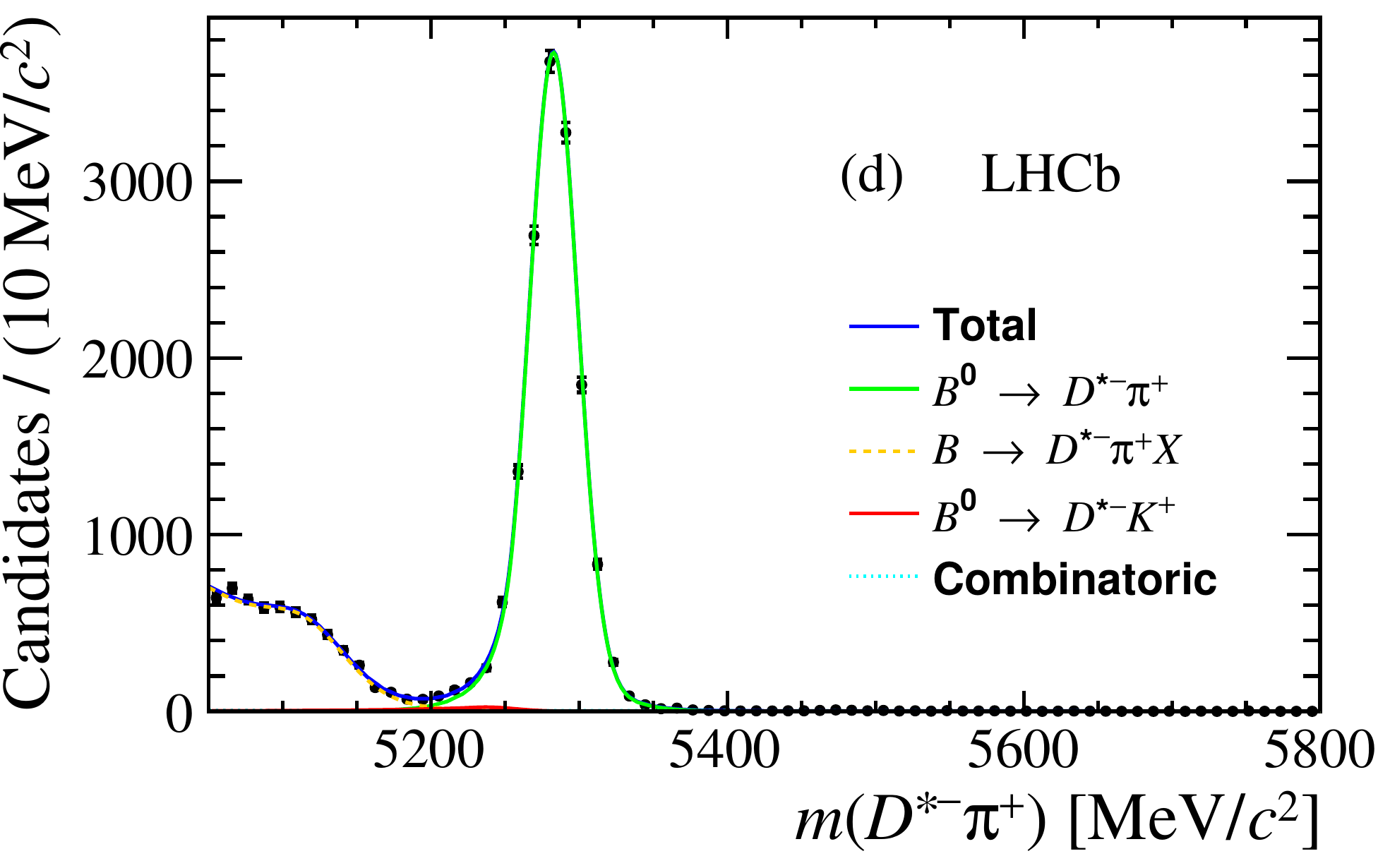}}\\
\caption{\small{Invariant mass distributions of selected \mbox{(a) $B^{0} \rightarrow D^{*-} K^{+} \pi^{-} \pi^{+}$}, \mbox{(b) $B^{0} \rightarrow D^{*-} K^{+}$},  \mbox{(c) $B^{0} \rightarrow D^{*-} \pi^{+}\pi^{-}\pi^{+}$} and \mbox{(d)  $B^{0} \rightarrow D^{*-} \pi^{+}$} candidates. Green (red) solid lines represent the CF (CS) signal shapes and their respective particle misidentification backgrounds. Orange dashed lines at low invariant mass represent backgrounds from partially reconstructed decays. Cyan dotted lines represent combinatorial backgrounds. The pink dash-dotted line below the signal peak in the fit to $B^{0}\rightarrow D^{*-}K^{+}\pi^{-}\pi^{+}$ candidates represents the background from misidentified $B^{0}\rightarrow D^{*-}D_{s}^{+}$ and $B^{0}\rightarrow D^{*-}K^{+}\Kstarzb$ decays.\label{fig:Fits}}}
\end{center}
\end{figure}

The reconstructed invariant mass distributions for $B^{0} \rightarrow D^{*-}h^{+}\pi^{-}\pi^{+}$ and $B^{0} \rightarrow D^{*-}h^{+}$ decays are shown in Fig.~\ref{fig:Fits}. Simultaneous binned maximum likelihood fits to the CF and CS decays are performed, where the probability density functions (PDFs) are composed of a signal component and several background components. The CF signal shapes are required to share parameters with their CS counterparts. The total CF (CS) signal yield $\mathcal{N}_{\text{CF}}$ ($\mathcal{N}_{\text{CS}}$) is given by the sum of the yield contained in the signal shape and the yield contained in the particle misidentification background from the CS (CF) sample. The values of $\mathcal{N}_{\text{CF}}$ and the ratio $\mathcal{N}_{\text{CS}}/\mathcal{N}_{\text{CF}}$ vary freely in the fit. 

The PDF of the signal decays is described by the sum of a Crystal Ball~\cite{Skwarnicki:1986xj} and a Gaussian function, when all final state particles are assigned the correct mass hypothesis. This shape was chosen as it describes radiative loss and the non-Gaussian mass resolution. In the simultaneous fit to the CF and CS decays, the Crystal Ball widths vary freely and independently for both modes. A single freely varying peak position parameter is shared by all signal components in the fit to both modes. The Gaussian width is required to be equal to or greater than the Crystal Ball width, and this width is shared by the CF and CS signal shapes.

All decay modes have background contributions from partially-reconstructed decays of the type $B \rightarrow D^{*} X$, where $X$ represents the final state bachelor(s) for the given decay plus an additional photon or pion that is not reconstructed. These backgrounds are parameterised by the sum of two single-sided Gaussian functions with a common mean and independent widths, all of which vary freely. A combinatorial background is present in all cases and is fit by an exponential function. The yields of the partially-reconstructed and combinatorial backgrounds vary freely in all parts of the fit to data.

Misidentification of pions and kaons causes cross-feed between the CF and CS signal decays. The simulation indicates that backgrounds of this type can be described by a PDF comprising two Crystal Ball functions that share a common mean but have independent widths. The absolute rate of a background from particle misidentification is fixed to be \mbox{($100 - \epsilon_{\text{PID}}$)\%} of the corresponding signal yield in the simultaneous fit, where $\epsilon_{\text{PID}}$ represents the efficiency for all final state hadrons in the signal decay to be correctly identified. For the single bachelor decays, $\epsilon_{\text{PID}}$ is determined by re-weighting the $\bigtriangleup LL(K-\pi)$ distributions obtained from calibration events to match the properties of the signal bachelor. The re-weighting is performed in bins of momentum, pseudorapidity and number of tracks in the event. Calibration tracks are taken from $D^{*+}\rightarrow D^{0}\pi_{s}^{+}$ decays with $D^{0}\rightarrow K^{-}\pi^{+}$~\cite{arXiv:1211-6759}. For the triple-bachelor decays a similar approach is used, but the kinematic correlations between each bachelor are also considered. A per-candidate particle identification efficiency is determined from the product of each bachelor PID efficiency, and $\epsilon_{\text{PID}}$ is given by the weighted average of the per-candidate efficiencies.

Further backgrounds are present in the $B^{0} \rightarrow D^{*-}K^{+}\pi^{-}\pi^{+}$ sample from the decays $B^{0} \rightarrow D^{*-}D_{s}^{+}$ where $D_{s}^{+} \rightarrow K^{+}K^{-}\pi^{+}$ and the $K^{-}$ meson is misidentified as a $\pi^{-}$, and \mbox{$B^{0} \rightarrow D^{*-}K^{+}\Kstarzb$} where $\Kstarzb \rightarrow K^{-}\pi^{+}$ and the $K^{-}$ is misidentified as a $\pi^{-}$. The backgrounds are modelled together using a Crystal Ball shape that peaks at a lower mass than the signal, with a peak position and width that vary freely in the fit. The Crystal Ball tail parameters are fixed to the values found in simulation. This background could be reduced by applying tighter particle identification requirements to the $\pi^{-}$ bachelor, but this has not been applied in order to maintain symmetry between the CF and CS particle identification requirements. To eliminate the background from $B^{0} \rightarrow D^{*-}D_{s}^{+}$ decays where $D_{s}^{+} \rightarrow K^{+}\pi^{-}\pi^{+}$, the veto previously described is applied. The background from $B^{0} \rightarrow D^{*-}D_{s}^{+}$ decays with $D_{s}^{+} \rightarrow \pi^{+}\pi^{-}\pi^{+}$ does not contribute due to the tight particle identification requirement applied to the bachelor kaon in the $B^{0} \rightarrow D^{*-}K^{+}\pi^{-}\pi^{+}$ decay. 

The fits are superimposed on the data in Fig.~\ref{fig:Fits}, and the measured yields $\mathcal{N}$ for each decay are listed in Table~\ref{tab:Fit_Yields}.  The CF and CS ratios of branching fractions are obtained using
\begin{align*}
r_{3h} &= f_{3h} \times \frac{\mathcal{N}(B^{0}\rightarrow D^{*-}\pi^{+}\pi^{-}\pi^{+})/\epsilon_{\text{tot}}(B^{0}\rightarrow D^{*-}\pi^{+}\pi^{-}\pi^{+})}{\mathcal{N}(B^{0}\rightarrow D^{*-}\pi^{+})/\epsilon_{tot}(B^{0}\rightarrow D^{*-}\pi^{+})}\, , \\
r_{\text{CS}}^{K\pi\pi} &= f_{\text{CS}}^{K\pi\pi} \times \frac{\mathcal{N}(B^{0}\rightarrow D^{*-}K^{+}\pi^{-}\pi^{+})/\epsilon_{\text{tot}}(B^{0}\rightarrow D^{*-}K^{+}\pi^{-}\pi^{+})}{\mathcal{N}(B^{0}\rightarrow D^{*-}\pi^{+}\pi^{-}\pi^{+})/\epsilon_{\text{tot}}(B^{0}\rightarrow D^{*-}\pi^{+}\pi^{-}\pi^{+})}\, , \\
r_{\text{CS}}^{K} &= f_{\text{CS}}^{K} \times \frac{\mathcal{N}(B^{0}\rightarrow D^{*-}K^{+})/\epsilon_{\text{tot}}(B^{0}\rightarrow D^{*-}K^{+})}{\mathcal{N}(B^{0}\rightarrow D^{*-}\pi^{+})/\epsilon_{tot}(B^{0}\rightarrow D^{*-}\pi^{+})}\,.
\end{align*}
The values of $\epsilon_{\text{tot}}$ are listed in Table~\ref{tab:Effs}, and the $f$ factors correct for systematic effects.

\begin{table}[tpb]
\begin{center}
\begin{tabular}{p{4cm}|ccc} 
Decay & \phantom{00}Yield \\ \hline
$B^{0} \rightarrow D^{*-} \pi^{+} \pi^{-} \pi^{+}$ & \phantom{0}7228 $\pm$ 93 \\ 
$B^{0} \rightarrow D^{*-} \pi^{+}$ & \phantom{,}15,693 $\pm$ 136 \\
$B^{0} \rightarrow D^{*-} K^{+} \pi^{-} \pi^{+}$ & \phantom{0}\phantom{0}519 $\pm$ 30 \\ 
$B^{0} \rightarrow D^{*-} K^{+}$ & \phantom{0}1241 $\pm$ 53\\ 
\end{tabular}

\caption{\small{Selected candidate yields from fits to data that are used in the branching fraction calculations. The yield for a decay is given by the sum of the signal shape yield in the CF (CS) fit and the corresponding misidentification background yield in the CS (CF) fit. Uncertainties quoted are statistical only.\label{tab:Fit_Yields}}}
\end{center}
\end{table}

\begin{table}[tpb]
\begin{center}
\begin{tabular}{p{4cm}|ccc} 
Decay & $\epsilon_{\text{kin}}$ (\%) & $f_{\text{trig}}$ (\%) & $\epsilon_{\text{tot}}$ (\%) \\ \hline
$B^{0} \rightarrow D^{*-} \pi^{+} \pi^{-} \pi^{+}$ & 0.037 $\pm$ 0.001 & 69.3 $\pm$ 0.5 & 0.0259 $\pm$ 0.0005  \\ 
$B^{0} \rightarrow D^{*-} \pi^{+}$ & 0.197 $\pm$ 0.002 & 75.4 $\pm$ 0.3 & 0.148 $\pm$ 0.002  \\
$B^{0} \rightarrow D^{*-} K^{+} \pi^{-} \pi^{+}$ & 0.044 $\pm$ 0.001 & 67.4 $\pm$ 0.9 & 0.0298 $\pm$ 0.0007  \\ 
$B^{0} \rightarrow D^{*-} K^{+}$ & 0.201 $\pm$ 0.003 & 75.4 $\pm$ 0.5 & 0.151 $\pm$ 0.002 \\     
\end{tabular}

\caption{\small{Kinematic efficiencies, trigger pass fractions and their product, taken from simulation. Quoted uncertainties come from the use of finite size samples to determine efficiencies and are accounted for as a source of systematic uncertainty.\label{tab:Effs}}}
\end{center}
\end{table}

\section{Systematic uncertainties}

By measuring ratios of branching fractions, many sources of systematic uncertainty cancel. The systematic uncertainties for the $r_{3h}$, $r_{\text{CS}}^{K\pi\pi}$ and $r_{\text{CS}}^{K}$ measurements will be discussed in turn. The primary sources of systematic uncertainty that remain in the $r_{3h}$ measurement are due to the different topologies of the signal and normalisation decays. Compared to the $B^{0}\rightarrow D^{*-}\pi^{+}$ normalisation mode, the triple-pion decay mode has two additional pions which must be reconstructed and selected. The tracking efficiency has been studied using a tag-and-probe method with $J/\psi \rightarrow \mu^{+} \mu^{-}$ decays~\cite{LHCb-PUB-2011-025}, which leads to a correction in $r_{3h}$ of 1.017 $\pm$ 0.035. In the $B^{0}\rightarrow D^{*-}\pi^{+}\pi^{-}\pi^{+}$ decay, the three bachelor pions are required to have a common vertex. The IP resolution and vertex $\chi^{2}/$ndf distributions are observed to be $\sim$15\% broader in data relative to the simulation~\cite{LHCb-PAPER-2011-016}, resulting in a correction on $r_{3h}$ of 0.982 $\pm$ 0.016.

Possible background from decays of the type \mbox{$B^{0} \rightarrow D^{*-}D_{s}^{+}$}, where \mbox{$D_{s}^{+} \rightarrow \pi^{+}\pi^{-}\pi^{+}$}, has been considered, and a correction of 0.990 $\pm$ 0.005 is applied to $r_{3h}$. The use of simulated events to determine trigger pass fractions has a residual systematic uncertainty arising from differences between data and simulation with respect to the emulation of the hardware trigger and trigger software. A correction of 1.009 $\pm$ 0.012 is applied to $r_{3h}$ to account for this difference. The candidate selection is limited to the mass region $m(\pip\pim\pip) <$ 3 GeV/$c^{2}$. The $m(\pip\pim\pip)$ distributions in data and simulation are in good agreement, such that the selection efficiency properly accounts for this choice. The fraction of events falling beyond 3 GeV/$c^{2}$ in simulation is 3.7\%. Assuming 50\% uncertainty on this value, a relative systematic uncertainty of 1.9\% is assigned.

The methods used to determine $\epsilon_{\text{PID}}$ have an uncertainty from which the systematic contribution is determined to be 0.8\%. A systematic uncertainty of 0.6\% arises from the specific choice of PDF shapes in the fit. Both of the CF simulated samples have a comparable number of events after selection requirements are imposed, from which a 2.1\% systematic uncertainty due to finite simulated samples is incurred.

The CF and CS $B^{0} \rightarrow D^{*-}h^{+}\pi^{-}\pi^{+}$ modes have identical selection requirements, apart from the particle identification requirements placed on the $h^{+}$ and the $D_{s}^{+}$ veto applied in the CS case. A systematic uncertainty of 3.4\% is incurred as a result of the particle identification requirement placed on the bachelor kaon in the $B^{0} \rightarrow D^{*-}K^{+}\pi^{-}\pi^{+}$ mode. This tight requirement is necessary in order to reduce the background from misidentified $B^{0} \rightarrow D^{*-}\pi^{+}\pi^{-}\pi^{+}$ decays. To evaluate the loss of signal events due to the $D_{s}^{+}$ veto in the CS selection, the fit to data is performed both with and without the veto applied. The measured CS signal yield decreases by 1\% upon application of the veto, which is taken as an inefficiency with 50\% uncertainty and a correction of 1.010 $\pm$ 0.005 is applied to $r_{\text{CS}}^{K\pi\pi}$.

The ratio of trigger pass fractions taken from simulation is \mbox{$f_{\text{trig}}(B^{0}\rightarrow D^{*-}\pi^{+}\pi^{-}\pi^{+})/f_{\text{trig}}(B^{0}\rightarrow D^{*-}K^{+}\pi^{-}\pi^{+}) =$ 1.03 $\pm$ 0.02}, where the quoted uncertainty is derived from the size of the simulated samples. For such similar decay modes, this ratio should be close to unity. The ratio itself is already applied as part of the branching fraction calculation, but half of the difference in the ratio from unity (1.5\%) is taken as an additional source of systematic uncertainty. 

In a similar fashion to the CF measurement, the CS measurement has a systematic uncertainty of 1\% from the specific choice of PDF shapes and 3.0\% uncertainty from the use of finite simulated samples to determine efficiencies. The fraction of CS decays with $m(K^{+}\pi^{-}\pi^{+})$ in the range 2.7--3 GeV/$c^{2}$ in data is 5.5\%. The fraction of events falling beyond the analysis cut at 3 GeV/$c^{2}$ is estimated to be half of this value (2.8\%) with 50\% uncertainty. The value of $r_{\text{CS}}^{K\pi\pi}$ is therefore corrected by a factor 1.028 $\pm$ 0.023, where the quoted uncertainty contains a 1.9\% contribution from the corresponding systematic uncertainty in the CF decay.  

The $B^{0} \rightarrow D^{*-}h^{+}$ modes have identical selection requirements apart from the particle identification requirements placed on the $h^{+}$ bachelor. A systematic uncertainty of 2.0\% is incurred as a result of the particle identification requirements applied to the bachelors. Further systematic uncertainties of 0.7\% , 1.7\% and 2.0\% arise from the specific choice of PDF shapes, the use of finite simulated samples and the trigger emulation~\cite{Aaij:1507868}, respectively.

Each contribution to the systematic uncertainty is listed in Table~\ref{tab:Systematics}. The total systematic uncertainty is given by the sum in quadrature of all contributions. The overall systematic uncertainty for the CF measurement is 5.0\%, with a factor \mbox{$f_{3h} =$ 0.998} that is applied as part of the calculation for $r_{3h}$. The CS triple- and single-bachelor measurements have overall systematic uncertainties of 5.4\% and 3.4\% respectively. A factor $f_{\text{CS}}^{K\pi\pi} =$ 1.038 is applied to $r_{\text{CS}}^{K\pi\pi}$, whereas $f_{\text{CS}}^{K} =$ 1.

\begin{table}[!t]
\begin{center}
\begin{tabular}{p{6cm} c c c} 
Source & \multicolumn{3}{c}{Value (\%)} \\
 & $r_{3h}$ & $r_{\text{CS}}^{K\pi\pi}$ & $r_{\text{CS}}^{K}$ \\ \hline
Track reconstruction & 3.4 & $-$ & $-$ \\
Selection requirements &  1.6 & $-$ & $-$ \\
$B^{0}\rightarrow D^{*-}D_{s}^{+}$ background & 0.5 & $-$ & $-$ \\
Trigger & 1.2 & 1.5 & 2.0\\
$m(h^{+}\pi^{-}\pi^{+}) >$ 3 GeV/$c^{2}$ & 1.9 & 2.2 & $-$\\ 
Particle identification & 0.8 & 3.4 & 2.0 \\
Choice of PDFs & 0.6 & 1.0 & 0.7 \\
Simulated sample size & 2.1 & 3.0 & 1.7 \\
$D_{s}^{+}\rightarrow K^{+}\pi^{-}\pi^{+}$ veto & $-$ & 0.5 & $-$\\ \hline
Total & 5.0 & 5.4 & 3.4 \\ 
\end{tabular}
\caption{Contributions to the relative systematic uncertainty for all measurements. The total uncertainty is obtained by adding the contibutions from the individual sources in
quadrature.\label{tab:Systematics}}
\end{center}
\end{table}

\section{Results}

The results for the ratios of branching fractions are
\begin{align*}
r_{3h} &= 2.64 \pm 0.04\,(\text{stat.}) \pm 0.13\,(\text{syst.})\, ,\\
r_{\text{CS}}^{K\pi\pi} &= (6.47 \pm 0.37\,(\text{stat.}) \pm 0.35\,(\text{syst.})) \times 10^{-2}\, ,\\
r_{\text{CS}}^{K} &= (7.76 \pm 0.34 \,(\text{stat.}) \pm 0.26 \,(\text{syst.})) \times 10^{-2}\, ,
\end{align*}
where the first uncertainty is statistical and the second is absolute systematic. Using the world average value for $\mathcal{B}(B^{0}\rightarrow D^{*-}\pi^{+}) = $ (2.76 $\pm$ 0.13) $\times$ 10$^{-3}$~\cite{PDG2012}, the branching fractions are obtained
\begin{align*}
\mathcal{B}(B^{0}\rightarrow D^{*-}\pi^{+}\pi^{-}\pi^{+}) &= (7.27 \pm 0.11\,(\text{stat.}) \pm 0.36\,(\text{syst.}) \pm 0.34\,(\text{norm.})) \times 10^{-3}\,,\\
\mathcal{B}(B^{0}\rightarrow D^{*-}K^{+}) &= (2.14 \pm 0.09 \,(\text{stat.}) \pm 0.08 \,(\text{syst.}) \pm 0.10 \,(\text{norm.})) \times 10^{-4} \,,
\end{align*}
where the final uncertainty is due to the normalisation mode. Both results are consistent with and improve upon the precision of the current world average values \mbox{$\mathcal{B}(B^{0}\rightarrow D^{*-}\pi^{+}\pi^{-}\pi^{+}) =$ (7.0 $\pm$ 0.8) $\times 10^{-3}$} and \mbox{$\mathcal{B}(B^{0}\rightarrow D^{*-}K^{+}) =$ (2.14 $\pm$ 0.16) $\times 10^{-4}$}~\cite{PDG2012}. Combining the CF result and the current world average, where both values are weighted according to their total uncertainty, gives \mbox{$\mathcal{B}(B^{0}\rightarrow D^{*-}\pi^{+}\pi^{-}\pi^{+}) =$ (7.19 $\pm$ 0.43) $\times 10^{-3}$}. The measurement of $r_{\text{CS}}^{K\pi\pi}$ represents a first observation of the decay \mbox{$B^{0}\rightarrow D^{*-}K^{+}\pi^{-}\pi^{+}$}. The value of this ratio is similar to the related measurement of \mbox{$\mathcal{B}(B^{0}\rightarrow D^{-}K^{+}\pi^{-}\pi^{+})/\mathcal{B}(B^{0}\rightarrow D^{-}\pi^{+}\pi^{-}\pi^{+}) =$ (5.9 $\pm$ 1.1\,(\text{stat.}) $\pm$ 0.5\,(\text{syst.})) $\times$ 10$^{-2}$}~\cite{LHCb-PAPER-2011-040}. Using the updated world average value for $\mathcal{B}(B^{0}\rightarrow D^{*-}\pi^{+}\pi^{-}\pi^{+})$, the branching fraction is obtained
\begin{equation*}
\mathcal{B}(B^{0}\rightarrow D^{*-}K^{+}\pi^{-}\pi^{+}) = (4.65 \pm 0.26\,(\text{stat.}) \pm 0.25\,(\text{syst.}) \pm 0.28\,(\text{norm.})) \times 10^{-4}\,.
\end{equation*}

\begin{figure}[tpb]
\begin{center}
  \includegraphics*[width=0.6\textwidth]{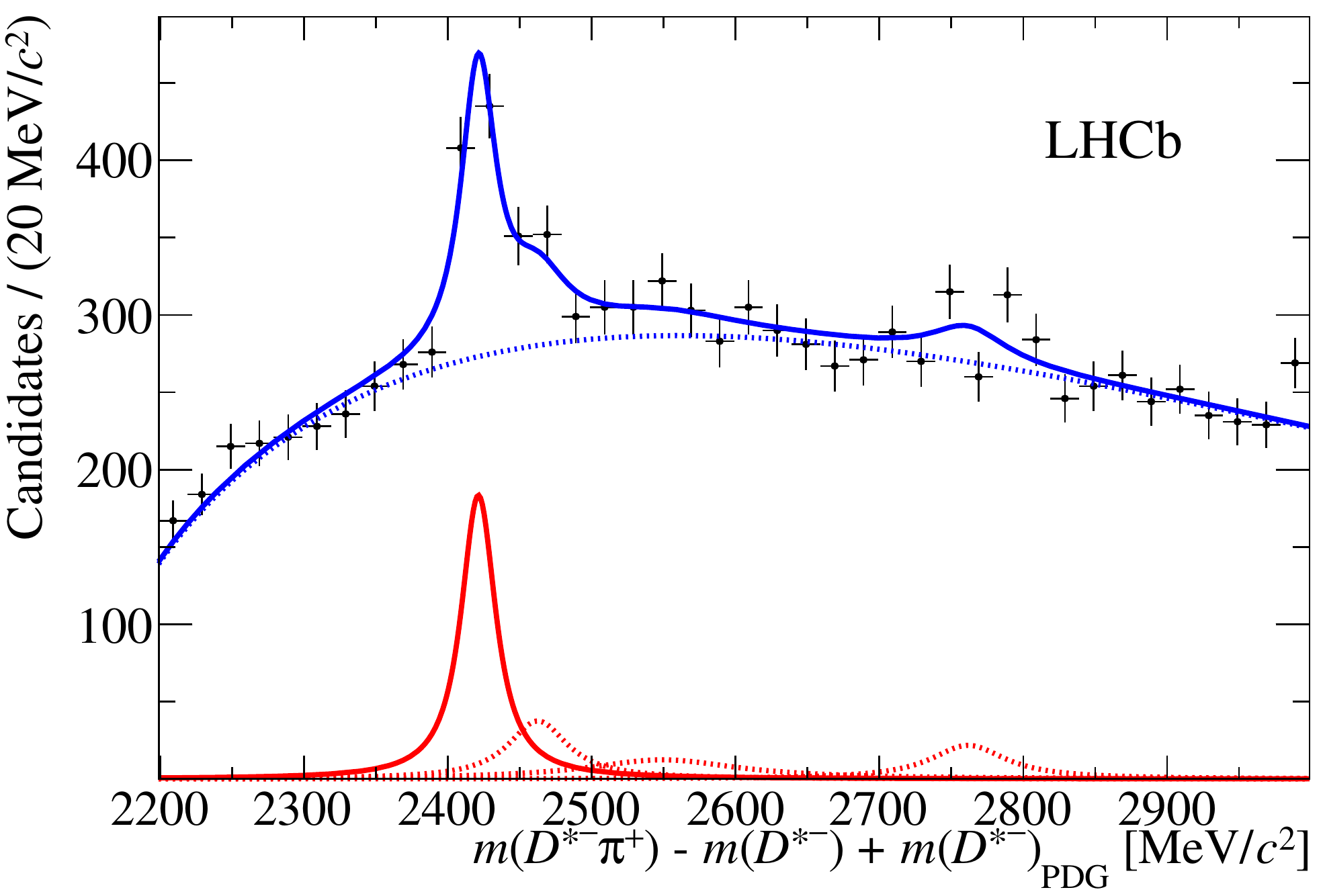}
\caption{\small{Corrected mass $M(D^{*-}\pi^{+}) = m(\Dstarm\pip)-m(\Dstarm)+m(\Dstarm)_{\text{PDG}}~\mevcc$ distribution from background-subtracted \mbox{$B^{0}\rightarrow D^{*-}\pi^{+}\pi^{-}\pi^{+}$} candidates. The red solid line represents the contribution from $\Dbar_1(2420)^0$, the red dashed lines represent contributions from the higher excited charm states $D^*_2(2460)^0$, $D(2550)^{0}$, $D(2600)$ and $D(2750)$, which are expected to be present~\cite{Dstarstar} but are not significant in this dataset. The blue dashed line represents $B^{0}$ decays that have not passed through an excited charm resonance. \label{fig:Dstarstar_Plots}}}
\end{center}
\end{figure}

\section{Search for excited charm resonances}

Using the same dataset, a search for orbital excitations of charm resonances ($D^{**}$) contributing to the $\Bz\to\Dstarm\pip\pim\pip$ final state is performed. The selection is identical to that presented above, except the lower $\pt$ cut on bachelors is reduced to 300 \mevc. Events that have been triggered via tracks not associated with the $\Bz\to\Dstarm\pip\pim\pip$ candidate are also included. The signal purity is only slightly reduced as a result of the looser selection.

The corrected mass \mbox{$M(D^{*-}\pi^{+}) =m(\Dstarm\pip)-m(\Dstarm)+m(\Dstarm)_{\text{PDG}}~\mevcc$} is computed for each $D^{*-}\pip$ combination, in which the contribution to the mass resolution from the $D^{*-}$ mass measurement is removed. To statistically subtract the background, each event is weighted using \textit{sWeights}~\cite{sPlot} obtained from the \mbox{$B^{0}\rightarrow D^{*-}\pi^{+}\pi^{-}\pi^{+}$} invariant mass fit. Fig.~\ref{fig:Dstarstar_Plots} shows the resulting distribution of $M(D^{*-}\pip)$ for $\Bz\to\Dstarm\pip\pim\pip$ signal decays.

A peaking structure associated to the $\Dbar_1(2420)^0$ resonance is observed, where $\sim$90\% of the candidates in the peaking structure originate from the combination with the softer $\pi^{+}$ meson. Other resonances, consistent with $D^*_2(2460)^0$, $D(2550)^{0}$, $D(2600)$ and $D(2750)$, are included in the fit but are not found to be significant. All resonances in the fit are described by Breit-Wigner functions. The $\Dbar_1(2420)^0$ Breit-Wigner function is convolved with a Gaussian resolution function of 3 MeV/$c^{2}$ width. The means and natural widths of all peaking structures vary around their established values~\cite{Dstarstar} with Gaussian constraints.

The background from $B^{0}\rightarrow D^{*-}\pi^{+}\pi^{-}\pi^{+}$ decays that do not pass through an excited charm resonance is described by a function comprising the two-body phase space equation multiplied by an exponential acceptance function, $e^{-\alpha M(D^{*-}\pip)}$. The shape parameter $\alpha$ and all yields vary freely. The branching fraction ratio is calculated by comparing the fitted $\Dbar_1(2420)^0\to\Dstarm\pip$ yield with the total number of accepted $D^{*-}\pi^{+}\pi^{-}\pi^{+}$ events in the sample. A correction is taken from simulation to account for the acceptance, reconstruction and selection efficiency for events in the region close to $m(\Dbar_1(2420)^0)$ relative to the efficiency averaged across the full phase space. The ratio of efficiencies is $f = 0.91 \pm 0.04$. A systematic uncertainty of 10\% is assigned to the choice of background PDF, which is determined by remeasuring the $\Dbar_1(2420)^0$ yield after shifting the fit range by $\pm50$~\mevcc. 

The measured yield is $\mathcal{N}(\Bz\to \Dbar_1(2420)^0\pip\pim) = 203 \pm 42$ and the total number of $B^{0}$ signal events after the looser selection is $\mathcal{N}(\Bz\to \Dstarm\pip\pim\pip) = 10,939 \pm 105$. Using these values, the ratio of branching fractions is obtained
\begin{equation*}
\frac{\mathcal{B}(B^{0}\rightarrow (\Dbar_{1}(2420)^{0} \to D^{*-} \pi^{+}) \pi^{-}\pi^{+})}{\mathcal{B}(B^{0} \rightarrow D^{*-}\pi^{+}\pi^{-}\pi^{+})} = (2.04 \pm 0.42\,(\text{stat.}) \pm 0.22\,(\text{syst.})) \times 10^{-2}\, .
\end{equation*}
where the numerator represents a product of the branching fractions \mbox{$\mathcal{B}(B^{0}\rightarrow \Dbar_{1}(2420)^{0} \pim \pip)$} and \mbox{$\mathcal{B}(\Dbar_{1}(2420)^{0} \to D^{*-} \pi^{+})$}. The Wilk's theorem statistical significance of the $\Dbar_1(2420)^0$ peak is 5.9$\sigma$, which becomes 5.3$\sigma$ when the systematic uncertainty is included. This constitutes the first observation of the colour-suppressed $\Bz\to \Dbar_1(2420)^0\pip\pim$ decay.\\

\section{Summary}

In conclusion, $B^{0}\to D^{*-}h^{+}\pi^{-}\pi^{+}$ decays have been studied using $B^{0}\to D^{*-}h^{+}$ decays for normalisation and verification. The branching fractions of $B^{0}\to D^{*-}\pi^{+}\pi^{-}\pi^{+}$ and $B^{0}\to D^{*-}K^{+}$ decays are measured, and the CS $B^{0}\to D^{*-}K^{+}\pi^{-}\pi^{+}$ and colour-suppressed $B^{0}\to \Dbar_{1}(2420)^{0}\pi^{+}\pi^{-}$ decays are observed. The final results are
\begin{equation*}
r_{3h} = 2.64 \pm 0.04\,(\text{stat.}) \pm 0.13\,(\text{syst.})\, ,\\
\label{eq:CF_BF_ratio_result}
\end{equation*}
\begin{equation*}
r_{\text{CS}}^{K\pi\pi} = (6.47 \pm 0.37\,(\text{stat.}) \pm 0.35\,(\text{syst.})) \times 10^{-2}\, ,\\
\label{eq:CS_BF_ratio_result}
\end{equation*}
\begin{equation*}
r_{\text{CS}}^{K} = (7.76 \pm 0.34\,(\text{stat.}) \pm 0.26\,(\text{syst.})) \times 10^{-2}\, ,\\
\label{eq:CS_BF_ratio_result}
\end{equation*}
\begin{equation*}
\frac{\mathcal{B}(B^{0}\rightarrow (\Dbar_{1}(2420)^{0} \to D^{*-} \pi^{+}) \pi^{-}\pi^{+})}{\mathcal{B}(B^{0} \rightarrow D^{*-}\pi^{+}\pi^{-}\pi^{+})} = (2.04 \pm 0.42\,(\text{stat.}) \pm 0.22\,(\text{syst.})) \times 10^{-2}\, .
\end{equation*}

The results for $r_{3h}$ and $r_{CS}^{K}$ represent an improvement in precision, and the measurements of the decays $B^{0} \rightarrow D^{*-}K^{+}\pim \pip$ and $B^{0}\rightarrow (\Dbar_{1}(2420)^{0} \to D^{*-} \pi^{+}) \pi^{-}\pi^{+}$ both constitute first observations.








\section*{Acknowledgements}

\noindent We express our gratitude to our colleagues in the CERN
accelerator departments for the excellent performance of the LHC. We
thank the technical and administrative staff at the LHCb
institutes. We acknowledge support from CERN and from the national
agencies: CAPES, CNPq, FAPERJ and FINEP (Brazil); NSFC (China);
CNRS/IN2P3 and Region Auvergne (France); BMBF, DFG, HGF and MPG
(Germany); SFI (Ireland); INFN (Italy); FOM and NWO (The Netherlands);
SCSR (Poland); ANCS/IFA (Romania); MinES, Rosatom, RFBR and NRC
``Kurchatov Institute'' (Russia); MinECo, XuntaGal and GENCAT (Spain);
SNSF and SER (Switzerland); NAS Ukraine (Ukraine); STFC (United
Kingdom); NSF (USA). We also acknowledge the support received from the
ERC under FP7. The Tier1 computing centres are supported by IN2P3
(France), KIT and BMBF (Germany), INFN (Italy), NWO and SURF (The
Netherlands), PIC (Spain), GridPP (United Kingdom). We are thankful
for the computing resources put at our disposal by Yandex LLC
(Russia), as well as to the communities behind the multiple open
source software packages that we depend on.



\addcontentsline{toc}{section}{References}
\bibliographystyle{LHCb}
\bibliography{main,LHCb-PAPER,LHCb-CONF}

\end{document}